# Subwavelength focusing of light by a tapered microtube

Jian Fu,[a)] Hongtao Dong, and Wei Fang

*State Key Laboratory of Modern Optical Instrumentation, Department of Optical Engineering,*

*Zhejiang University, Hangzhou 310027, China*

[a)]Electronic mail: jianfu@zju.edu.cn

We propose a mechanism for subwavelength focusing at optical frequencies based on the use of a tapered microtube fabricated from a glass capillary tube. Using coherent illumination at 671nm and a near-field scanning optical microscope probe which was mounted on a 3-axis piezo nanopositioning stage, a sequence of 2-D intensity profiles were obtained. Our experimental results reveal the smallest focal spot with a near diffraction-limited full width at half-maximum of ~435nm（0.65λ）at a distance of ~1.47$\mu$m (2.2λ) from the output endface of microtube. The experimental results are in excellent agreement with our numerical simulation.

In recent years, there has been an explosion of interest in obtaining subwavelength focusing that can reach beyond the Abbe-Rayleigh diffraction limit, which remains a dominant barrier to achieving features smaller than approximately half a wavelength with optical instruments. Higher optical resolution will be beneficial to such fields as semiconductor lithography, optical trapping[1] and high density optical data storage.[2] Several approaches have been proposed and experimentally realized to meet these challenges, and the most successful ones are: 1. Plasmonic lenses based on surface plasmon polaritons (SPP), such as nanoslits, nanoholes and surface corrugations in a metallic film;[3-6] 2. So-called superlens, a perfect lens first proposed by Pendry, making use of materials or structures exhibiting negative refraction;[7-10] 3. Non-plasmonic lenses via Talbot effect,[11] in the form of nanohole quasi-periodic array in a metal screen,[12] a circle of planar nanoholes in a dielectric film[13] and a micro/nanofiber (MNF) array.[14] However, the focusing effects are still severely influenced by the metallic absorption for plasmonic lenses, and the superlenses are only capable of projecting an image with perfect resolution in the near field in experiments so far, due to the intrinsic losses.[15] While non-plasmonic lenses made from dielectric materials have considerably lower loss than the previous two types.

The concept of direct synthesis of the angular spectrum has been proposed asserting that a focused beam can be synthesized directly from its angular spectrum of plane waves.[16] And，in Ref. 16, the feasibility of this approach for focusing of light with subwavelength resolution has also been experimentally verified by using a designed apparatus converting 15 laser beams to constitute a converging circular cone. Inspired by this intriguing approach, in this letter, we propose a mechanism for subwavelength focusing waveguide of the tapered microtube where the cone part at optical frequencies by using a tapered microtube. As shown in Fig. 1, a laser beam propagates through sidewall converts the input laser beam into a circular cone. With this rather simple structure, we demonstrate that the laser beam focal spot size is close to diffraction limit, which agree well with our numerical simulation results.

The microtube investigated in this work was fabricated by flame-heated drawing method. A glass capillary tube with outside diameter and sidewall thickness as 2.0mm and 0.15mm, respectively, was heated by alcohol burner flame, and drawn to form a tapered microtube. Scanning electron micrograph (SEM) images of a typical fabricated tapered microtube are shown in Fig. 2 (a) and (b). As we can see, the output endface shows a fairly flat and smooth ring-shape cross section, with outside diameter of 4.8$\mu$m and sidewall thickness of 412nm. Figure 2 (c) shows an optical microscope image of a tapered microtube with a half opening angle $\alpha$ (taper ratio) of ~18° at the cone section. As we will discuss later, taper ratio is an important factor that will affect the final focal spot size.

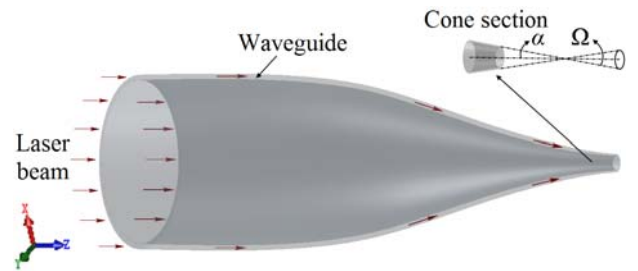

FIG. 1. (Color online) Schematic of a tapered microtube with a cone section and propagation of waves through the sidewall waveguide of the microtube. $\alpha$ and $\Omega$ denote, respectively, half angle and solid angle corresponding to the taper ratio of the cone section.

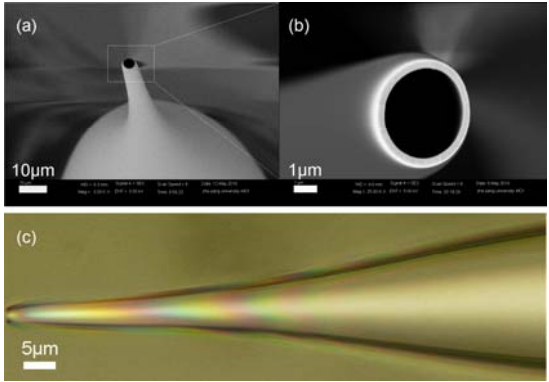
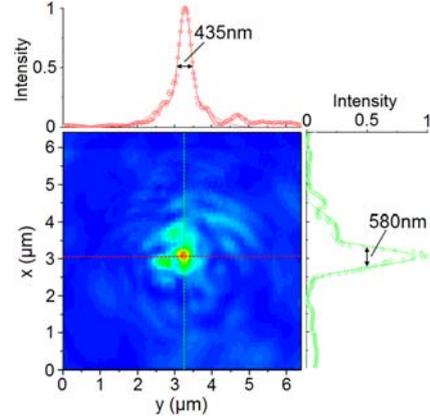

FIG. 2. (Color online) Images of a taped microtube. (a) SEM images of a well-fabricated microtube. (b) Magnified image of the output endface: a fairly flat and smooth ring-shape cross section, ~4.8$\mu m$ outside diameter of the ring and ~412nm sidewall thickness. (c) Optical microscopy (100×) image of a well-fabricated microtube with a half angle (taper ratio) of ~18°.

FIG. 3. (Color online) A fine scan 2-D intensity profile with the smallest spot at a distance of ~1.47$\mu m$ from the output endface of the microtube and corresponding profiles along the perpendicular green dash line and red dash line across through the center of the spot, with a FWHM of ~580nm and ~435nm, respectively.

The tapered microtube that was used in our optical measurement had the outside diameter of the ring of 2.7$\mu m$, sidewall thickness of 353nm at the output endface, and half opening angle $\alpha$ as 9°. The microtube was fixed on an optical stage with its axis oriented along the z direction. And a laser beam of 5mW at a wavelength of 671nm was first converged by a lens and then coupled into the input endface along the z direction. The refractive index of the microtube is 1.45 at this wavelength. A near-field scanning optical microscope (NSOM) probe (NT-MDT™, MF002), with a transmission coefficient of $1\times10^{-4}$, was mounted on a 3-axis piezo nanopositioning stage (PI™, P-282.20 PZT Flexure Stage) with its axis oriented along the z direction. A 3-channel high voltage piezoelectric transducer amplifier (PI™, E-463.00) was used to drive this stage, and the output voltage of each channel ranges from 0 to -1500V. The output signal from the probe was measured by a photomultiplier tube (PMT) with a digital multimeter (Agilent™ 34410), and stored and processed by a computer. During the experiments, the NSOM probe was scanning in the x-y plane at certain distance away from the output endface of the microtube. The optical intensity distribution was then mapped over a grid of 200×200 points spanning the scanning area of 6.4×6.4 $\mu m^2$, with 32nm/pixel resolution. Because the piezo nanopositioning stage we used was operated at the open-loop travel mode, nonlinear errors between forward and back displacements in the form of hysteresis phenomena occurred in the scan process. After correcting these nonlinear errors, we finally obtained the optical intensity profile of the output laser beam from the tapered microtube.

Figure 3 shows an optical intensity profile obtained at a distance of ~1.47$\mu m$ away from the output endface of the microtube. Besides several small side spots, we find one sharp spot at the center. By extracting the intensity reading cross the maximum point along y direction, we plotted the normalized intensity distribution, shown as the red curve on the top of Fig. 3. The full width at half-maximum (FWHM) of the peak is measured as 435nm, which is smaller than the wavelength of the incident laser. While extracting the intensity distribution along the perpendicular direction, we got a FWHM as 580nm, which indicated an elliptical shape of the focal spot. Taking the aperture size of the NSOM tip into account, we deconvoluted the distribution with a rectangular window function, and finally calculated the actual size of the spot estimated to be about 560×410$nm^2$ (0.83$\lambda$×0.61$\lambda$).

In order to investigate the intensity profile of the transmitted beam along the propagation direction, we numerically calculated the axis intensity profile by using 3-dimension beam propagation method (3D-BPM). The parameters of microtube were set the same as the one investigated in the experimental measurements. Figure 4 (a) shows the intensity distribution along the cross section of x-z plane. The output endface of the microtube (showed in the white dash lines) was located at z = 0 $\mu m$ plane. The propagating mode of the microtube at 671nm was launched as the input field. As we can find, there exists no clear focal spot in the region from z = 0$\mu m$ to z = 1.4$\mu m$, but a series of concentric doughnut-shaped circles. While in the region from z = 1.4$\mu m$ to z = 2.9$\mu m$, there shows an elongated focusing beam along the axis.

To verify these results, we repeated the same scanning process, and obtained a sequence of 2-D intensity profiles at z = 1$\mu m$, 1.94$\mu m$, and 2.41$\mu m$ planes. As shown in Fig.4 (b) − (e), the intensity distributions represent the main features of the numerical simulation results. Figure 4 (b) shows the intensity profile at z = 1$\mu m$ plane, with only several faint ring structures. In the region where z > 1.4$\mu m$, however, a bright spot emerges at the axial position, as shown in Fig. 4 (c) − (e). At z = 1.47$\mu m$, we find the smallest focal spot size; while at z=1.94$\mu m$, the focal spot has the maximum intensity over all four scans, but larger FWHM than the one at z = 1.47$\mu m$. At z = 2.41$\mu m$, we find even larger FWHM of the focal spot, with relatively smaller intensity. All of these observations fit well with numerical simulation results shown on Fig. 4 (a).

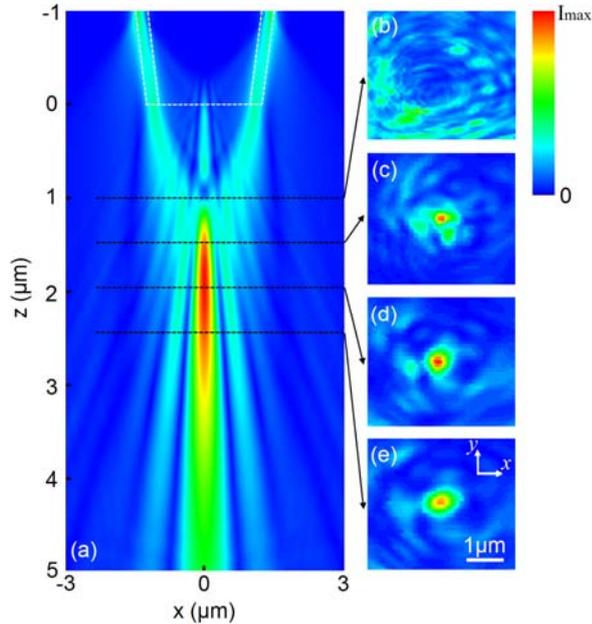

FIG. 4. (Color online) (a) Calculated axial intensity profile (on the x-z plane) of the simulation results using 3-dimension beam propagation method. (b)-(e) Measured 2-D intensity profiles (on the x-y plane) of (a) at z=1, 1.47, 1.94, and 2.41$\mu$m, respectively.

As we have demonstrated above, both the measurements and the simulation show clearly subwavelength focusing of the input laser. The reasons that we can achieve such tight focusing have been elaborated in Ref. 16, namely, direct synthesis of the angular spectrum. Different from their discrete 15 laser beams setup, we utilized tapered microtube to simulate the ideal Bessel beam with continuous function. Analogous to the corresponding intensity distribution $I(r)$ suggested in Ref. 16, the spot intensity here depends on the available wave vectors of the interfering waves, that is, on the solid angle $\Omega$ from which the interfering waves area comes, as depicted in Fig. 1. The microtube with the larger solid angle will have a higher power and produce a tighter focal spot, although it will come with the price of shortening the focal length of the device. In consequence, the taper ratio of the cone section plays a critical role in dominating the focusing performance of the microtube. With properly designed structure parameters, we shall be able to get even better results.

Unlike the planar plasmonic lenses that subject to high losses due to the inherent imaginary part of the metal's permittivity, the microtube described in this letter is a purely dielectric system and therefore has a low intrinsic material loss. The extrinsic losses that are expected in this system only result from the imperfections in the fabrication, coupling losses and output endface losses. According to the rules suggested in Ref. 13, a higher rotational symmetry leads to a higher focusing intensity in the foci; the ring with the highest rotational symmetry has pretty high energy efficiency in focusing. Moreover, in Ref. 11 and Ref. 12, the foci created by quasi-periodic array of nanoholes in a metal screen are such sparsely distributed that each focal spot failed to harvest energy from all the holes in the array. The advantage of the microtube in this respect is that it produces only one focal spot. The present device might find applications in areas such as semiconductor lithography, optical trapping, and optical data storage.

In conclusion, we have demonstrated, experimentally and numerically, the subwavelength focusing by using a tapered microtube. Single focal spot has been observed, with FWHM as narrow as 0.65$\lambda$ at a distance of several wavelengths. Using dielectric waveguide structure, we are able to achieve high transmission power, which results in high optical intensity at focal spot.

This work was supported by the National Natural Science Foundation of China under Grant No.60407003, and National Basic Research Program 973 of China under Grant 2007CB307003. The authors thank Professors Xu Liu and Li-Min Tong for helpful discussion.